# Classical Teleportation of Classical States

Oliver Cohen
6 Bigpath Farm Cottages, Upham, Hampshire, SO32 1HL, UK[1]

The standard quantum teleportation scheme is deconstructed, and those aspects of it that appear remarkable and "non-classical" are identified. An alternative teleportation scheme, involving only classical states and classical information, is then formulated, and it is shown that the classical scheme reproduces all of these remarkable aspects, including those that had seemed non-classical. This leads to a re-examination of quantum teleportation, which suggests that its significance depends on the interpretation of quantum states.

The process of *quantum teleportation* was first described in the seminal paper of Bennett et al. [1]. In the ten years since its publication, it has had a huge influence on the field of quantum information, triggering a race for the first experimental implementation [2, 3], and generating hundreds of journal citations. Quantum teleportation has also captured the imagination of the general public, with descriptions in the popular science literature that typically evoke images from the TV series Star Trek. More recently, there has been worldwide headline coverage of the successful teleportation of a laser beam [4].

At first sight, the extent of the interest and publicity that quantum teleportation has given rise to is not difficult to understand. The instantaneous transfer across an arbitrary spatial distance of a quantum state that can be defined to arbitrary precision, a state that is completely unknown to both sender and receiver, seems startling, remarkable, and, in particular, *non-classical*. Hence it may seem quite natural that teleportation should be considered a key part of the repertoire of quantum information processes.

In the following, our aim is first of all to break down the standard quantum teleportation protocol into a number of steps, and to identify those aspects of the procedure that appear to capture the essence of the non-classicality of the technique. We will then describe a very simple method for teleportation of a type of purely *classical* information, using classical systems. Surprisingly, and notwithstanding what has been said so far, this classical technique is very closely analogous to the standard quantum teleportation protocol. Indeed, it transpires that each of the particular aspects of quantum teleportation that we identified as appearing to display non-classicality is reproduced exactly in the purely classical version of the technique! This leads us to re-examine the significance of quantum teleportation. Our conclusion is that the question of whether or not quantum teleportation has any special significance depends on our interpretation of quantum states. If quantum states are interpreted (as is commonly the case) as representing *states of knowledge*, then it is hard to see that quantum teleportation achieves anything important beyond what can be achieved with a purely classical process. It is only if quantum states are given some kind of ontological status that quantum teleportation represents a process that can never be mimicked classically.

---

[1] e-mail: oliver@bigglebaggle.com

We start with a description of quantum teleportation as outlined in [1] (We also include some comments which are intended to facilitate comparison with the purely classical scheme, a description of which is to follow):

(1) An EPR pair is distributed between Alice and Bob.
(2) Charlie prepares a two-level quantum system in the state $|\psi\rangle$ and gives it to Alice. (Comments: (i) Although the state $|\psi\rangle$ represents a 2-dimensional system, it requires a doubly infinite amount of information to describe it, because it is defined by two continuously variable parameters. (ii) It is impossible to *measure* $|\psi\rangle$, given just a single system is that state.).
(3) Alice carries out a joint measurement on the system that Charlie has just given her and her component of the EPR pair that was originally distributed between her and Bob. This measurement has four possible outcomes. (Comments: (i) as soon as Alice has completed her measurement, she knows that the state of Bob's system will now be the same state $|\psi\rangle$ that Charlie supplied to her, up to a known unitary transformation, although she does not know what $|\psi\rangle$ is. In other words, from Alice's perspective, the teleportation process, up to a known unitary transformation, takes place instantaneously. (ii) After this measurement, Alice has no information whatsoever about the state of either *individual* system in her possession, but only has information about the *joint* state of both systems.)
(4) According to which result is obtained in the measurement described in (3), Alice sends 2 bits of classical information to Bob.
(5) Having received the 2 bits of information from Alice, Bob performs one of four possible operations on his system. (Comment: One of these operations is the identity operation, i.e. it consists of doing nothing.)
(6) Following the operation referred to in (5), the state of Bob's system will be $|\psi\rangle$, i.e. it will be an exact copy of the original state supplied by Charlie to Alice.

At this point it is worth pausing to highlight those aspects of the above process that seem remarkable and non-classical:

(a) The state $|\psi\rangle$ requires a doubly infinite amount of information to define it, yet it is teleported from Alice to Bob with the aid of only two bits of information.
(b) The state $|\psi\rangle$ is teleported faithfully from Alice to Bob without either party at any time having any information about the state whatsoever.
(c) The kernel of the teleportation process, as seen by Alice, takes place instantaneously.

(d) All trace of the original state $|\psi\rangle$ at Alice's location is erased in the teleportation process.

We now give an overview of our 100% *classical* teleportation protocol; we will then go on to give a more detailed explanation of the classical protocol. This initial overview is intended to provide a preliminary comparison with the quantum teleportation protocol as described above.

(1) A correlated pair of two-level classical systems is distributed between Alice and Bob. (The particular type of correlation, and the method of preparation, will be explained shortly.)

(2) Charlie prepares a two-level classical system in the state $|x)$. We will explain what is meant by "the state $|x)$" presently; suffice to say at this stage that the concept of the state $|x)$, and its realization, are purely classical. (Comments: (i) Although the state $|x)$ represents a two-dimensional system, it requires an infinite amount of information to describe it, because it is defined by one continuously variable parameter. (ii) It is impossible to *measure* $|x)$, given just a single system in that state.)

(3) Alice carries out a joint measurement on the system that Charlie has just given her and her component of the correlated pair that was originally distributed between her and Bob. This measurement has two possible outcomes. (Comments: (i) As soon as Alice has completed her measurement, she knows that the state of Bob's system will now be the same state $|x)$ that Charlie supplied to her, up to a known simple transformation, although she does not know what $|x)$ is. In other words, from Alice's perspective, the teleportation process, up to a known simple transformation, takes place instantaneously. (ii) After this measurement, Alice has no information whatsoever about the original state of either *individual* system in her possession, she does however have information about the *joint* state of both systems.)

(4) According to which of the two possible outcomes is obtained in the measurement described in (3), Alice sends 1 bit of classical information to Bob.

(5) Having received the bit of classical information from Alice, Bob performs one of two possible operations on his system. (Comment: one of these two possible operations is the identity operation, i.e. it consists of doing nothing.)

(6) Following the operation as described in (5), the state of Bob's system will be $|x)$, i.e. it will be an exact copy of the original state supplied by Charlie to Alice.

Before making a comparison of the steps involved in the two respective teleportation protocols, and examining the extent to which the remarkable aspects (a) – (d) of the quantum protocol, as highlighted above, are retained in the classical case, we will explain what we mean by "the state $|x)$", and describe how the classical technique could be implemented.

By "the state $|x)$" we simply mean the following:

A classical system is in one of two possible configurations. One of these possible configurations is the reference configuration. Suppose we do not know which configuration the system is in, but we know that the probability that it is in the reference configuration is x. Then we say that the state of the system is $|x)$.

For example, suppose that we are given an ensemble of N coins, of which Nx are showing heads face up and N(1-x) are showing tails. Suppose that we assign "heads face up" as our reference configuration. If we then select a coin from the ensemble, but do not look to see whether it is showing heads or tails face up, then our description of the coin will be $|x)$.

Note that the notion of a classical state $|x)$, as we have defined it here, is epistemological – it refers to a *state of knowledge* with respect to an observer. However, this is not really very different to the way that many people think about quantum states – that is, that they are a way of encapsulating our knowledge about a system, rather than having any direct ontological interpretation [5].

We now describe the method we will use for preparing a classical state $|x)$ for the purposes of our classical teleportation protocol. We begin by preparing N ensembles of coins in boxes, where N is large. Each of the N ensembles consists of N coins, each of which is placed inside its own individual box, which is then sealed. The coin is placed in its box showing either heads or tails face up, before the box is sealed. The boxes are perfectly symmetrical, and each coin fits its box tightly, so that it cannot move within its box once the box is sealed. Hence, if a coin is placed in its box showing heads face up, and the box is rotated through 180 degrees, then the system of box plus coin will be identical to one where the coin was placed in its box showing tails face up and the box was not rotated. Once again we assign "heads face up" (within a box this time) as our reference configuration.

We can label these N ensembles 1 through N. In ensemble 1, one coin is placed in its box showing heads face up, and (N-1) showing tails. In ensemble 2, two coins are placed in their boxes showing heads face up, and (N-2) showing tails, and so on; i.e., in ensemble n, n coins are placed in their boxes showing heads face up and (N-n) showing tails.

To prepare a coin in any chosen state $|x)$, we simply select a sealed box from the ensemble numbered Nx. (Regardless of the accuracy to which x is expressed, we can ensure that Nx is a whole number by making N sufficiently large.) If the box is chosen at random from this ensemble, then we know that the probability that the coin inside it is showing heads is x, hence our description of the coin will be $|x)$. (Note that we can convert the state $|x)$ to the state $|1-x)$ by turning the box upside down.)

It is possible to prepare a pair of coins-in-boxes in a correlated state – for example, a state where we know that, with probability y, both coins are showing heads face up within their boxes, and, with probability (1-y), both coins are showing tails face up, i.e. there is zero probability of the two coins showing opposite faces. We denote such a state $|y\rangle_{HH} + |1-y\rangle_{TT}$. Such a state could be straightforwardly prepared using a method similar to that described above, but this time with Ensemble n containing N pairs of boxes of which n pairs contain coins both of which are showing heads face up within their boxes, and (N-n) pairs contain coins which are both showing tails face up. To prepare the state $|y\rangle_{HH} + |1-y\rangle_{TT}$ we would then select a pair of boxes from Ensemble Ny.

We are now in a position to explain how the classical teleportation scheme can be implemented:

(1) A pair of coins-in-boxes is prepared in the state $|1/2\rangle_{HH} + |1/2\rangle_{TT}$, and the boxes are then distributed to Alice and Bob at their respective locations. The state $|1/2\rangle_{HH} + |1/2\rangle_{TT}$ could be prepared using the ensemble method described above. Alternatively, a simple way of preparing this state using just a single pair of coins-in-boxes (rather than N ensembles of N pairs) is as follows: Two coins are placed in boxes showing heads face up and the boxes are then sealed. The boxes are then both rotated through the same random number of 180 degree turns. The rotation could be carried out by a machine coupled to a random number generator, or we could recruit an external protagonist to rotate the two boxes through any number of 180 degree turns they choose and then return them to us. Following the rotation, we will know that both coins are showing the same face within their respective boxes, but we will have no information whatsoever as to whether that face is heads or tails. Hence our description of the pair will be $|1/2\rangle_{HH} + |1/2\rangle_{TT}$. Once this state has been prepared, the boxes must be distributed to Alice and Bob whilst ensuring that no further rotations take place.

(2) Charlie prepares a single coin-in-a-box in the chosen state $|x\rangle$, using the ensemble method as described earlier. Charlie then passes the box over to Alice.

(3) Alice carries out a joint measurement on the box that was originally distributed to her as part of the correlated pair, and the box that Charlie has just given her, as follows: She arranges that both her boxes are rotated by the same random number of 180 turns. Once again this rotation can be carried out by a

(3) ...machine or by an external protagonist. Following the rotation the boxes are opened (by the machine or the external protagonist) and the coins examined to see if they are showing the same (i.e. both heads or both tails) or different (i.e. one heads and one tails) faces, and the outcome is relayed to Alice as "same" or "different".

(4) Alice sends one bit of information to Bob, according to whether the outcome of the measurement described in (3) is "same" or "different".

(5) If Bob receives the message "same", he leaves his box as it is; if he receives the message "different" he rotates his box by 180 degrees.

(6) Following the operation described in (5), the state of Bob's coin must be $|x)$. This can be seen straightforwardly as follows: The whole sequence of operations just described ensures that the coin inside Bob's box is showing the same face up as the one that was showing in Charlie's box at the time that he (Charlie) originally selected it. When he selects his box, Charlie knows that the probability that the coin inside it is showing heads is x; hence, the probability that Bob's coin is showing heads at the end of the process is x, i.e. the state of Bob's coin is $|x)$.

At no point in the process do Alice or Bob have any knowledge of $|x)$. The state $|x)$, as teleported from Alice to Bob, represents a description of the system *as given by Charlie*. If we compare this to the quantum teleportation protocol we can see that, unless we attach some ontological significance to a quantum state $|\psi\rangle$, then there is an exact correspondence in the sense that, in the quantum case, the teleported state $|\psi\rangle$ represents *Charlie's description of the system*. Indeed, in the quantum case, since Alice and Bob have no information whatsoever about the teleported state, their best final description of the system to which it is teleported is the maximally mixed state.

Now that we have explained how our classical teleportation protocol works, we can re-visit the remarkable and "non-classical" aspects of quantum teleportation that we identified earlier, to see to what extent, if any, these aspects are also present in the classical version. Recall that we earlier highlighted four remarkable features as follows:

(a) The state requires a (doubly) infinite amount of information to describe it, yet it is teleported from Alice to Bob with the aid of only two bits of information. *This also holds in the classical version, the only differences being that an infinite rather than doubly infinite amount of information is required to describe the classical state, and one rather than two bits of information must be sent from Alice to Bob. If we carry out the classical version twice, using two different states, we can match the quantum case, in that the two states will require in total a doubly infinite amount of information to describe them, and Alice will have to send a total of two bits of information to Bob.*

(b) The state is transferred faithfully from Alice to Bob without either party at any time having any information about the state whatsoever.
*This also holds in the classical version.*
(c) The kernel of the teleportation process, as seen by Alice, takes place instantaneously.
*This also holds in the classical version.*
(d) All trace of the original state at Alice's location is erased in the course of the teleportation process.
*This also holds in the classical version.*

Surprisingly, we can see that the "remarkable features" of the quantum teleportation process – the ones that had appeared to capture its uniqueness and *non-classicality* – remain present, in their entirety, in our purely classical version of the process!

What are we to conclude from this? Although quantum states are very different to classical states [6], it seems that our classical version of teleportation is just as impressive as the original quantum protocol, *if we think of quantum states as representing states of knowledge*. In both cases, the end result of the process is that Charlie's final description of Bob's system is identical to his (Charlie's) original description of his own system. From the epistemological standpoint, it is the status of being the object of Charlie's state of knowledge that is transferred from Alice's system to Bob's. If, on the other hand, we think of a quantum state as having ontological content, representing perhaps a real physical field [7], then our classical version of teleportation is not equivalent to the quantum case, because the latter now involves the transfer of intrinsic physical content from Alice's system to Bob's.

This interpretation-dependence of the significance of quantum teleportation is in contradistinction to other facets of quantum information processing, such as quantum computation and quantum cryptography, the significance of which have nothing to do with the interpretation of quantum states. This is because the significance of computation and cryptography lies in their practical usefulness, which is independent of any interpretational standpoint. Whilst it is true that quantum teleportation may have limited practical use, for example to enhance the range for quantum cryptography implementations, our analysis indicates that its fundamental significance as a stand-alone technique cannot be assessed without recourse to the question of interpretation.

Acknowledgement
I am grateful to Todd Brun for comments on the manuscript.

[6] For example, if a quantum system is in a pure quantum state $|\psi\rangle$, then we can always identify a measurement, the outcome of which, if carried out on the system, can be predicted with certainty. This is not the case for the classical state $|x)$ (which in this sense has more in common with a mixed quantum state such as $x|0\rangle\langle 0| + (1-x)|1\rangle\langle 1|$).

[7] See for example D. Bohm and B. J. Hiley, *The Undivided Universe: An Ontological Interpretation of Quantum Mechanics* (Routledge, London, 1993).